\documentclass[12pt,preprint]{aastex}

\usepackage{graphics,graphicx}

\begin{document}


\title{CHANDRA OBSERVATION OF THE MAGELLANIC CLOUD SUPERNOVA REMNANT 
0454$-$672 IN N9 }

\author{F. D. Seward}
\affil{Smithsonian Astrophysical Observatory, 60 Garden St., 
Cambridge MA 02138}

\author{R. M. Williams, Y.-H. Chu, J. R. Dickel\altaffilmark{1}}
\affil{University of Illinois at Urbana-Champaign, 1002 W. Green St.,
  Urbana,IL 61801}
\altaffiltext{1}{Dept. of Physics and Astronomy, Univ. of New Mexico, 
800 Yale Blvd. NE, Albuquerque NM 87131}

\author{R. C. Smith and S. D. Points}
\affil{Cerro Tololo Inter-American Observatory, Casilla 603, 
La Serena, Chile} 

\begin{abstract}
A {\it Chandra} observation has defined the extent of the 
SNR\,B\,0454-692 in the LMC  \ion{H}{2} region N9.  The remnant has 
dimension $2\farcm3 \times 3\farcm6$  and is elongated in the NS 
direction.  The brightest emission comes from a NS central ridge 
which includes three bright patches.  There is good agreement between 
X-ray and [\ion{O}{3}] and [\ion{S}{2}] morphology. The remnant is old 
enough so that optical data give more information about dynamics than 
do the X-ray data. The SN energy release was $\geq 4 \times 10^{50}$ 
ergs and the age is $\sim 3 \times 10^{4}$ years.  There are several
unresolved sources nearby but none are clearly associated with the 
remnant.  The X-ray spectrum is soft and indicates enhanced Fe 
abundance in the central region, consistent with a Type Ia SN 
origin, but a Type II origin cannot be ruled out. 

\end{abstract}

\section{Introduction}

The supernova remnants (SNRs) within a galaxy enable studies
of the interaction between these objects and the interstellar 
medium (ISM), as well as often providing insight into the 
supernovae (SNe) that produced them.  About 50 remnants have 
been identified to date in Large and Small Magellanic Clouds 
(LMC, SMC). These remnants are all at known distances, which 
overcomes the bane of research using observations of Galactic 
remnants --- the distance uncertainty.  Distances of many 
Galactic remnants are uncertain by at least a factor of 2, leading 
to a factor of 4 uncertainty in luminosity, and to a factor of 5.5 
in the calculated energy release of the SN. Furthermore, the column 
density of hydrogen, $N_{\rm H}$, is low in the direction of the
Magellanic Clouds; thus soft X-ray emission, absorbed in the ISM and
consequently unseen from distant Galactic remnants, can be observed.  

The {\it Chandra} Observatory is well suited for observations of
SNRs in the  Magellanic Clouds.  Structure within the brighter 
remnants is well-resolved with moderate observing times and much 
detail is visible.  The {\it Chandra} arc-second resolution is 
particularly vital in the search for faint unresolved central 
objects. About 45 of the Magellanic Cloud remnants are known to 
emit X-rays \citep{S02,W+00,IKT83,F+98}.  So far, 30 of these 
have been, or are scheduled to be, observed by {\it Chandra}.  The 
small, bright remnants were observed first and observers are now 
working on larger, less luminous specimens.  SNR B 0454$-$672, the 
subject of this paper, is one of these.   These older remnants,
fading to X-ray invisibility, are an interesting population in 
themselves, as this stage takes up a large fraction of the life 
of a typical remnant. 

Several authors have worked with the LMC sample.  \citet{M+83} 
analyzed {\it Einstein} and optical data and concluded that the 
distributions of size and luminosity indicate more of a free-expansion 
than a blast-wave evolution.  \citet{CK88} have classified remnants 
through observations of the immediate environment. \citet{H+98} 
used spectral fits to derive abundances for seven middle-aged LMC 
SNRs, and thereby estimated mean ISM abundances in the LMC. \citet{H+95}
classified 6 LMC remnants as originating in a Type I or Type II SN
based on the strength of emission lines in the X-ray spectrum.  
\citet{HBR03} have similarly classified 2 remnants as SN Ia. 

SNR B 0454$-$692 was discovered by \citet{S+94} with the {\it ROSAT} 
PSPC detector, and confirmed through its  [\ion{S}{2}]/H$\alpha$ 
ratio and the presence of nonthermal radio emission.  The {\it ROSAT} 
X-ray data determined the size and X-ray luminosity of the SNR, and 
showed a center-filled morphology which often indicates the presence 
of a pulsar and associated pulsar-wind nebula (PWN).  The remnant is
immediately to the SW of N9, an \ion{H}{2} complex in the LMC.  The 
principle feature of N9 is a 10\arcmin\ diameter semicircular ring of 
H$\alpha$ emission which almost overlies the remnant. 

\section{Observations}

\subsection{X-ray Observations}

{\it Chandra} observed N9 for 68 ks on 2004 February 6 and 7 (Observation
ID 3847). Data were reprocessed using the \textsc{CIAO} software 
(version 3.2.2) provided by the {\it Chandra} X-ray Center (CXC). We 
applied corrections for charge-transfer inefficiency (CTI) and 
time-dependent gain for an instrument temperature of $\sim$120 C. We 
also used the 5$\times$5 pixel islands of our VFAINT mode observation 
for background cleaning. The data were filtered for poor event grades 
and restricted to the energy range of 0.3-8.0 keV, where the S3 chip 
is most sensitive. High charged-particle-induced background times noted 
in the pipeline processing were removed, as well as an additional 
3 ks high-background interval, resulting in a total ``good time"
exposure of $\sim$60 ks.

Using this filtered event file, $\sim$12,700 counts were obtained 
in the vicinity of the SNR, or about 9800 background-subtracted counts. 
Individual spectra for regions of interest, and the corresponding 
auxiliary response files, were extracted  from this event file with 
the \textsc{ciao} acisspec script; primary response files were 
separately generated with the mkacisrmf tool. Background regions 
were taken from source-free areas surrounding the SNRs, and these
background spectra were scaled and subtracted from the source 
spectra. Spectra were rebinned by spectral energy to achieve a signal-to-noise ratio of 3 in each bin. Subsequent analysis was
done using the \textsc{xspec} software.

\subsection{Optical Emission-Line Images}

Optical emission-line images of N9 were taken on 1997 December 1 
using the UM/CTIO Heber D. Curtis Schmidt telescope at Cerro Tololo 
Inter-American Observatory (CTIO) as part of the Magellanic Cloud 
Emission Line Survey (MCELS).  The detector was a thinned, back-side
illuminated Tek 2048 $\times$ 2048 CCD with 24 $\mu$m pixels, giving 
a 1\fdg3 $\times$ 1\fdg3 field-of-view with a scale of 2\farcs3 
pixel$^{-1}$ and a resulting angular resolution of $\sim$2\farcs6.  
The narrow band images were taken with filters centered on the 
[\ion{O}{3}] (5007 \AA, FWHM = 40 \AA), H$\alpha$ (6563 \AA, 
FWHM = 30 \AA), and [\ion{S}{2}] (6724 \AA, FWHM = 50 \AA) emission 
lines. 

We obtained ``continuum" band images of the SNR using green (5130 \AA, 
FWHM = 155 \AA) and red (6850 \AA, FWHM = 95 \AA) filters to allow for the 
subtraction of non-emission line sources (e.g., stars).  These observations 
were obtained in order to more accurately determine emission line fluxes as 
well as map out deeper detail in the faint, diffuse emission.  Two slightly 
offset images were obtained through each filter to allow for cosmic ray 
rejection and bad pixel replacement.  The total integration times were
1200 s in the [\ion{O}{3}] and [\ion{S}{2}] images and 600 s in the
H$\alpha$, green, and red images.

The data were reduced using the \textsc{IRAF} software package for 
bias subtraction and flat-field correction.  Astrometric solutions were
derived based on the {\it HST} Guide Star and USNO-A catalogs using 
automated WCS routines based on code from Brian Schmidt and \textsc{IRAF}
routines.  At this time, the images were also recast to have 2\arcsec
$\times$ 2\arcsec\ pixels. The images were background-subtracted using 
the peak of the histogram of pixel values as the best estimate of the 
sky.  The individual frames for each object were then aligned using the
astrometric solutions and multiple exposures in each filter were combined.

To isolate the line emission from continuum flux in the images, the 
continuum images were then scaled to the emission line images 
(normalizing by the ratio of counts in stars in the images) and 
subtracted.  Flux calibration was determined from observations of
spectrophotometric standards on many nights throughout the MCELS 
survey, and should be accurate to within 10\%.

\subsection{Radio Observations}

Radio data come from two sources: an 843 MHz image from the Molonglo
Observatory Synthesis Telescope (MOST; McIntyre \& Green 2000, personal
communication) and a 4.8 GHz image from a survey \citep{D+05} with the
Australia Telescope Compact Array (ATCA) combined with a survey using 
the Parkes Telescope \citep{H+91} to include both fine scale and extended
emission.

The MOST is an east-west array of two long co-linear paraboloids that 
can synthesize a target in a 12-hour observation.  The half-power beam 
width at 843 MHz is $43\arcsec \times 46\arcsec\ $ at position 
angle 0\degr. The image has an rms noise of about 1 mJy beam$^{-1}$.

The 4.8 GHz image was made from a survey of several quick looks at 
different hour angles and two different configurations of the five inner 
telescopes of the ATCA.  The synthesized half-power beam width is
33\arcsec.  Because of the limited data, the rms noise level 
on the images is about 0.3 mJy beam$^{-1}$. Polarimetric information was
recorded but the intensity of N9 was too faint for detection of any
polarization above the noise level.


Although the resolution and sensitivity are limited, we have measured the
integrated flux density, $S_f$ of the SNR, using the X-ray boundaries, at 
each wavelength.  We find 67 mJy at 843 MHz and 19 mJy at 4.8 GHz.  The 
errors in each are probably about 20\% due to uncertainties in separating 
the SNR from the \ion{H}{2} region and background plus the relatively 
weak signals.  The results give a power-law spectral index, $\alpha$ of
$-$0.7$\pm$0.2 (where $S_f \propto f^{\alpha}$).  This value is
characteristic of the synchrotron radiation from shell-type SNRs of
young age (Dickel 1991) but we cannot distinguish the progenitor type 
from this information.  We note that the \ion{H}{2} region has flux 
densities (within the H$\alpha$ outline) of 175 mJy at 843 MHz and 
215 mJy at 4.8 GHz.  Given the errors, these values are consistent with 
a flat spectral index characteristic of the thermal emission from 
\ion{H}{2} regions.

\section{Morphology}

\begin{figure}
\plotone{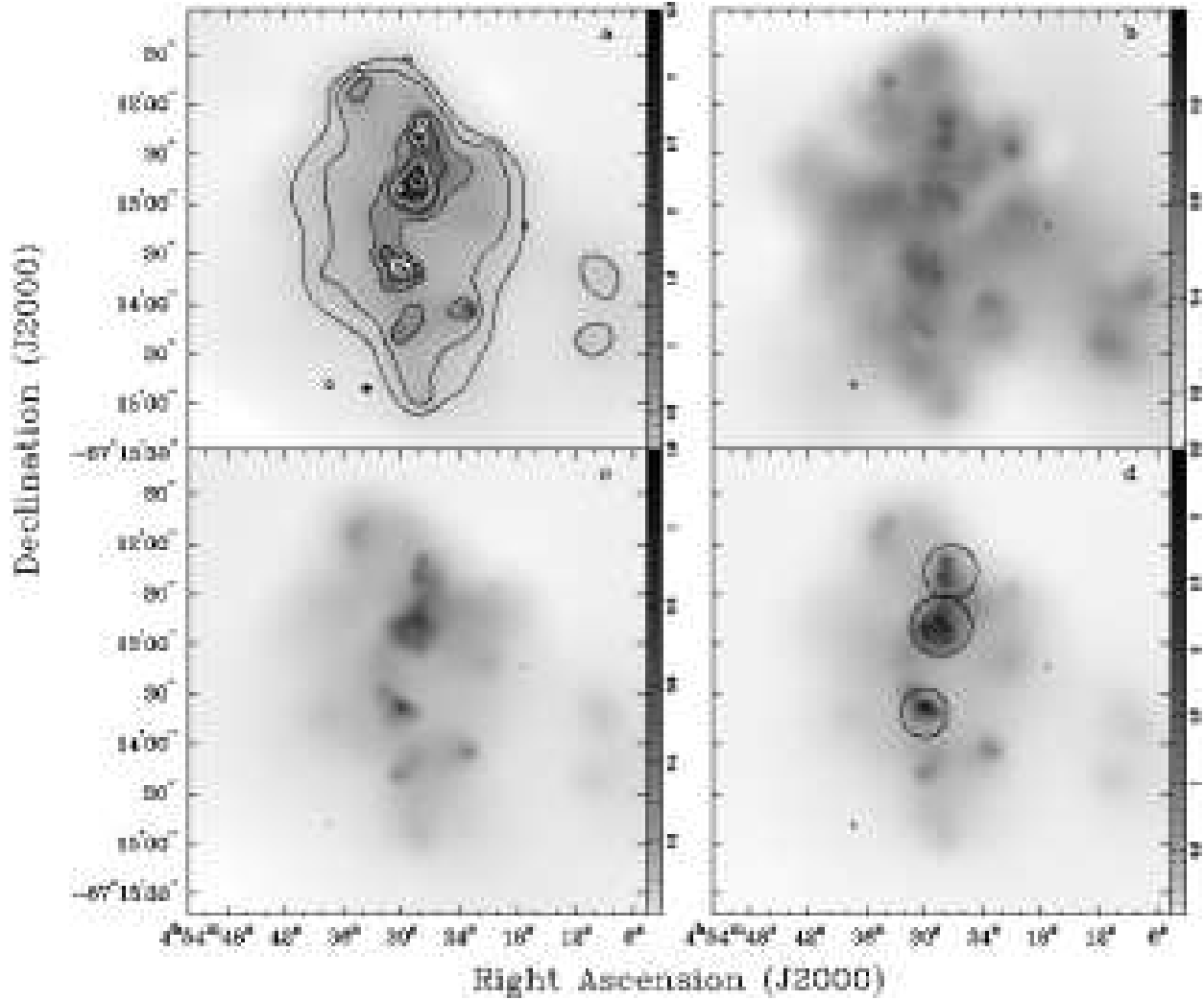}
\caption{(a) Adaptively smoothed {\it Chandra} ACIS image, with surface
brightness contours. Contours are logarithmically scaled over a factor
of 20 from faintest to brightest emission.  (b) Hardness ratio map 
H-S/H+S, using the soft and hard images below. (c) Soft-band  ACIS 
image showing emission at 0.3-0.7 keV. (d) Hard-band  ACIS image 
showing emission at 0.7-1.5 keV; regions used for the ``bright ridge"
spectrum are marked. The soft and hard X-ray images were adaptively 
smoothed on the same scale.}
\label{fig:xray}
\end{figure}

Figure~\ref{fig:xray}a shows an image in the energy range 
0.3--2.0 keV which contains almost all X-ray emission from the 
remnant. It has been adaptively smoothed with an algorithm that 
recognizes features with significance above 3$\sigma$. Contours 
of constant X-ray surface brightness are superposed on the image.  
Figure~\ref{fig:xray}b shows a map of the hardness ratio, defined 
as H-S/H+S, where S is is emission from 0.3--0.7 keV 
(Figure~\ref{fig:xray}c ) and H is emission from 0.7--1.5 keV
(Figure~\ref{fig:xray}d).  The low dynamic range in the hardness
ratio map demonstrates the spectral homogeneity across the SNR.

The remnant has dimension $2\farcm3 \times 3\farcm6$  and is 
elongated in the NS direction.  The brightest X-ray emission comes 
from a NS central ridge which includes three bright patches.  30\% 
of the emission comes from these 3 regions. The brightest of these 
patches is almost central and 0\farcm4 in diameter.  The surface 
brightness here is 5--10 times that of emission from the outer parts 
of the remnant.  There is a hint of limb brightening in the north 
but otherwise little to indicate an X-ray emitting shell.  

The optical morphology of SNR 0454$-$692 is highly filamentary 
\citep[for first publication of MCELS images, see][]{S+94}.
The filaments are distributed over a roughly elliptical region,
with distortions toward the western side of the remnant.
The H$\alpha$ emission from the SNR is superposed on a diffuse
component possibly associated with the nearby \ion{H}{2} region.
The filamentary structure appears clearer in the [\ion{S}{2}]
$\lambda\lambda$6716, 6731 image, showing the cooled post-shock
material.  The [\ion{O}{3}] emission outlines the rim of the SNR, 
more clearly delineating the expansion. The [\ion{S}{2}] 
filaments appear slightly inward of the [\ion{O}{3}]. This is 
expected for an expanding shell, as [\ion{O}{3}] is produced in
a temperature and density range that peaks closer to the shock 
front than the ranges for [\ion{S}{2}] or H$\alpha$.

\begin{figure}
\plotone{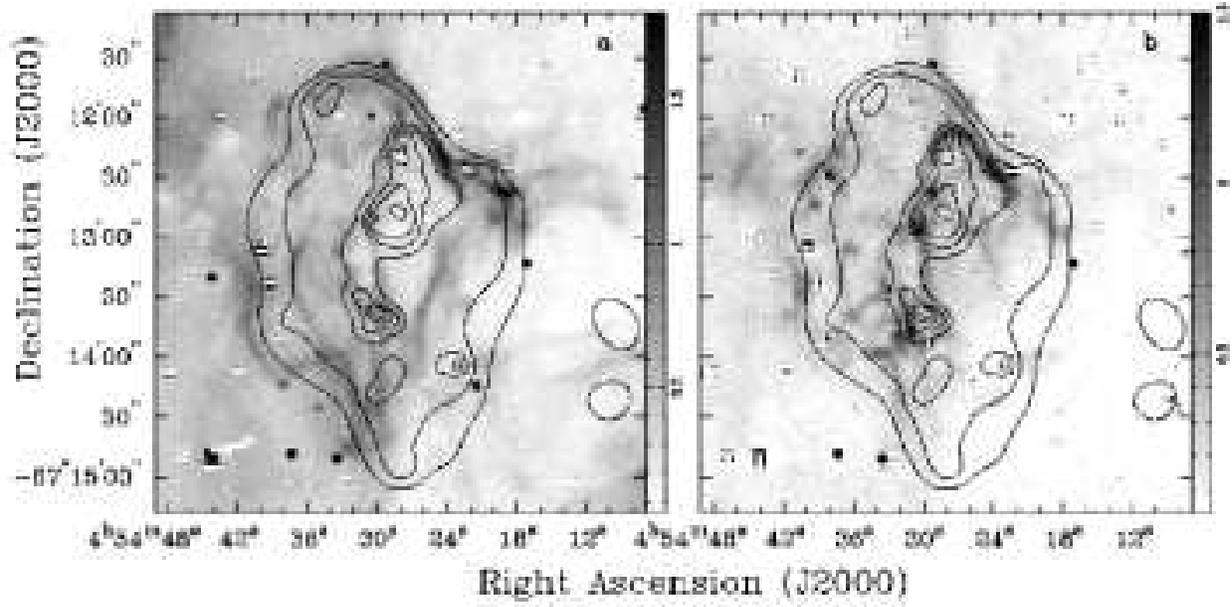}
\caption{(a) Adaptively smoothed ACIS X-ray contours overlaid
on a greyscale MCELS [\ion{O}{3}] image. (b) ACIS contours 
overlaid on a greyscale MCELS [\ion{S}{2}] image. Contours 
are scaled as in Fig.~\ref{fig:xray}.}
\label{fig:opt}
\end{figure}

Figure~\ref{fig:opt}a shows the X-ray contours overlaid on 
[\ion{O}{3}] emission from the SNR, which nicely follows the X-ray 
boundary in the NW and SE.   Figure~\ref{fig:opt}b shows an overlay 
on [\ion{S}{2}] emission.  The [\ion{S}{2}] emission follows the 
central ridge of brighter X-ray emission more strongly than the 
[\ion{O}{3}] emission.  To the SW, the X-rays appear to extend 
beyond the brightest optical emission.  Figure~\ref{fig:lgrad}a 
also displays the X-ray contours on [\ion{O}{3}] emission, this 
time on a larger scale, to show the relationship of the SNR to
the larger N9 region.  Figure~\ref{fig:lgrad}b shows X-ray 
contours overlaid on the 843 MHz radio emission for the same region.  
Although the radio resolution is limited, it can be seen that the 
basic structure of the radio emission corresponds well with that 
of the X-ray emission.  The brightest part of the \ion{H}{2} region 
to the northeast is also shown.  The SNR is detectable at 4.8 GHz 
but is faint and hard to separate from the \ion{H}{2} region. 

The X-ray images also show a patch of very faint emission to the west,
but no optical or radio counterpart to this emission is seen. Lacking 
any optical or radio evidence, we have chosen not to include this 
region as part of the remnant. However, it is certainly possible that 
this emission is in fact an extension of the SNR with connecting 
emission obscured by the dust lane seen in Figure~\ref{fig:opt}a. 
Better radio data are needed to give a definite answer.

\section{X-ray Spectra}

Spectral fits were made to two regions: the three bright patches in 
the NS ridge and the faint diffuse structure filling the remainder 
of the remnant.  (The three bright patches used for the central ridge 
are shown in Figure~\ref{fig:xray}d.)  A uniform distribution of material 
was assumed for each region.  Although the advanced age of the SNR (see 
\S 6.1) suggests it may have reached ionization equilibrium (CIE), the 
X-ray spectra of other remnants have shown nonequilibrium ionization 
(NEI) effects to late SNR ages.  We therefore fit the spectra with 
both NEI {\em vpshock} and CIE {\em vmekal} spectral models.  For 
both models, abundances of all elements except H and He were set at 
0.3 solar \citep[typical for the LMC;][]{RD92} and abundances of
O, Ne, and Fe allowed to vary.  The ISM column density was
fixed at $2 \times 10^{21}$ atoms cm$^{-2}$ \citep[Dunne 2004, 
personal communication, extracted from the  \ion{H}{1} survey 
of][]{KS+03}.  Results are listed in Table~\ref{tab:spec}. 

Figure~\ref{fig:spec} shows the best fits to 
the spectra for the central ridge and the faint diffuse regions.  The 
only significant difference in the spectra of the two regions is that 
the bright region has relatively more events in the range 0.65-0.9 keV 
than does the faint.  This is where L-shell emission from Fe is 
concentrated and the difference is accounted for by adjusting the 
relative abundance of Fe.  Abundances of O and Mg can be varied 
by up to 0.1 without making much difference to the fits.  The 
temperature, however, is strongly dependent on the column density 
which was not allowed to vary. 

The low ionization timescales ($\tau$) found for the NEI fits are
somewhat surprising for a SNR of the dynamical age estimated in 
\S 6.1.  If we freeze the timescale at a value consistent with CIE, 
and compare this fit to a fit in which the timescale is allowed to 
vary, an F-test between the two fit statistics shows the improvement 
to be significant. (The probability of no improvement is
$1\times10^{-11}$.)

To investigate variations between the volume emission measures of the 
three bright patches, the spectra were extracted separately for each 
of these regions, and then fit jointly with the best-fit model for the
spectrum of the ridge as a whole.  Only the normalizations, which are
proportional to the volume emission measures ($\int n_e n_H dV$), were
allowed to vary.  The resulting volume emission measures are given in 
Table~\ref{tab:xvolem}.  Also given are estimated volumes occupied
by the hot gas for each region. The bright patches are approximated
by spherical volumes; for the northern region, we also did a volume
estimate presuming that this region is part of the SNR shell.
The faint diffuse emission is approximated by an ellipsoidal region 
with the line-of-sight dimension set equal to the minor axis, minus 
the volumes of the bright patches.

\section{Compact Objects}

The presence of interior neutron stars and their related PWN
shows that gravitational collapse is the cause of some SN
explosions.   The association of a remnant and compact object
also links both ages which can be difficult to evaluate separately.
Many such associations should help us understand the evolution of
these objects. 

The LMC has been a fertile hunting ground.  Three LMC remnants are 
known, through detection of pulsations, to have internal neutron 
stars: SNR 0540$-$69.3 and N~157B with Crab-like central objects 
and pulsars \citep{K+01,W+01}, and N~49 with an SGR considerably 
off center \citep{M+79,E+80,R+94}.  PWNe have been detected in two
other SNRs, suggesting the presence of as yet undetected compact 
objects.  SNR 0453$-$685 has a PWN at the center, but no unresolved 
source which might be the pulsar is detected \citep{G+03}. N~206 
(0532$-$710) shows a faint X-ray shell surrounding bright thermal 
emission from the center.  There is a weak compact X-ray source 
and PWN at one end of a linear radio feature, close to the rim of 
the remnant.  \citet{W+05} propose this to be a fast moving neutron 
star leaving a radio wake. 

The three neutron stars from which pulsations are seen are bright, 
with $L_{X}\approx 10^{36}$ erg s$^{-1}$.  Fainter objects, however, 
would be harder to find. For example, there is a radio-quiet neutron 
star, or CCO for Central Compact Object, in the Galactic remnant 
Kes 79 \citep{S+03} with $L_{X} = 3\times 10^{33}$ erg s$^{-1}$ and a 
blackbody spectrum with kT $\sim0.4$ keV, very similar to the central 
source in Cas A.  \citep[A 105 ms period has recently been detected in 
this CCO by][showing it to be an X-ray pulsar.]{G+05}
At a distance of 7 kpc, this CCO produced 700 counts in a 30 ks 
{\it Chandra} observation.  Based on the greater distance but lesser
column density of the LMC, we estimate that a 60 ks observation of 
this object in the LMC would yield 90 counts.  So, in the LMC, a 
$10^{33}$ erg s$^{-1}$ CCO will yield 30 counts.  

On the other hand, the Vela Pulsar, the faintest ``young'' pulsar  
with $L_{X}\approx 2\times 10^{32}$ erg s$^{-1}$, would be undetectable 
in the LMC.  The absorption column density toward the LMC is an order 
of magnitude higher than that towards the Vela Pulsar, and thus the 
soft spectrum from such a pulsar would be absorbed.  The surrounding 
PWN, however, has $L_{X}\approx 10^{33}$ erg s$^{-1}$ and a hard 
spectrum.  It would have dimension of $0\farcs5$ and would be seen, 
at the 30 count level, as an unresolved source.  Thus, within the 
Magellanic Cloud remnants, we can easily detect central sources as 
faint as those found in the Milky Way. The upper limit for a point
source imbedded in the bright center of this remnant is $\approx 
5 \times 10^{32}$ erg s$^{-1}.$ 

The remnant under consideration here, 0454$-$672, shows no evidence 
for an energetic compact object.  The morphology of the central 
bright region suggests a PWN but the X-ray spectrum from this region 
is similar to that of the rest of the remnant. In contrast, the PWN 
in 0453$-$685, which is younger, shows a distinctly nonthermal 
spectrum at both radio and X-ray wavelengths \citep{G+03}.  There 
are seven nearby unresolved sources which are listed in 
Table~\ref{tab:src}.  None are central to the remnant but, in an 
older remnant such as this, an initial kick could have moved the 
object to or beyond the boundary.  The fourth source in the
table, just inside the western boundary, is the best candidate
for a CCO.  Although neither intensity nor spectrum distinguishes 
it from a typical field source, a CCO similar to that found in 
Kes 79 is not ruled out.  The sixth source in the table has a hard 
spectrum but is $3.3^{\prime}$ west of the remnant center, rather 
far away to have been associated with the SN.

\section{Remnant Structure}

\subsection{Cool Ionized Shell}

We divided the flux-calibrated [\ion{S}{2}] by H$\alpha$ to produce a [\ion{S}{2}]/H$\alpha$ ratio map (Figure~\ref{fig:s2ha}).  This map 
shows an irregular  inner region of filaments with [\ion{S}{2}]/H$\alpha$ 
ratios of 0.6--0.8, with an outer ``ring" of filaments with lower
[\ion{S}{2}]/H$\alpha$ ratios of 0.5--0.7, falling to 0.4 at the
boundary between the SNR and the \ion{H}{2} region.  These ratios are 
very similar to those measured by \citet[][Fig. 5,8d]{S+94} for this 
SNR; as observed in that paper, these measured ratios are somewhat 
lower than the true ratios, due to contaminating emission from
the background \ion{H}{2} region.

To determine the density of the warm ionized shell of the N9 SNR,
we have measured the average H$\alpha$ surface brightness of filaments
to be (1--2) $\times 10^{-15}$ erg s$^{-1}$ cm$^{-2}$ arcsec$^{-2}$.
The H$\alpha$ surface brightness can be converted to emission measure,
$EM = \int n_{\rm e}^2 dL$, where $n_{\rm e}$ is the electron density
and $L$ is the length of the emitting material.
Using the recombination coefficient for a 10$^4$ K gas \citep{O89} 
and assuming a filament's length along the line of sight to be
comparable to its length perpendicular to the line of sight,
44\arcsec\ $\pm$ 8\arcsec, we find an rms electron density of
$n_{\rm e} = 8 \pm 2$ cm$^{-3}$ in the SNR.

In the warm ionized gas, we assume He is singly ionized with a He 
to H number ratio of 0.1, and that the temperature of the gas is 
$\sim10^4$ K.  We can then use the electron density to calculate the 
thermal pressure in the shell: $ P_{\rm th} = 2.0 n_{\rm e} kT =
  2.3 \pm 0.4 \times 10^{-11}$ dyne cm$^{-2}$. Similarly, 
we infer the mass of gas in the shell from our calculated electron 
density: $ M_{\rm shl} = 1.27 n_{\rm e} m_{\rm H} V_{\rm shl}$,
where $V_{\rm shl}$ is the shell volume. We assumed that the bulk of 
the warm ionized gas is distributed in a shell with thickness equal to 
the measured width of the filaments (1.0 $\pm$ 0.3 pc).  Using this as
the shell volume, we calculate the mass in such a shell to be 
$1300 \pm 800 M_{\sun}$. From high-dispersion echelle spectra of the 
H$\alpha$ and [\ion{N}{2}] $\lambda\lambda$6548, 6583 lines, 
\citet{C97} estimated an average expansion velocity for the N9 SNR 
of $\sim 115 \pm 5$ km s$^{-1}$.  From this expansion velocity and 
our calculated shell mass, we determine a kinetic energy in the shell 
($0.5 M_{\rm shl} v_{\rm exp}^2$) of 1.7 $\pm 0.8 \times 10^{50}$ erg. 

We can also use this optical expansion velocity to estimate the 
age of the N9 SNR, using the simple expansion relationship 
$t = \eta R/v_{\rm exp}$, where $t$ is the SNR age and $R$ is 
its current radius.  If we assume blast-wave expansion 
\citep[$\eta=0.4$;][]{C72,S59} for this remnant, that velocity 
implies an age of 37,000 years; if we instead assume a pressure-driven snowplow model \citep[$\eta=0.33$;][]{BP04}, appropriate to the large 
size and interior X-ray emission, the calculated age drops to 31,000 
years.

\subsection{Hot Gas}

The density of X-ray emitting material was calculated for three 
regions: the brightest central region - 17\arcsec\ or 4.5 pc in 
radius; the faint diffuse interior; and the maximum in the north 
which might be a small part of the shell.  These densities were 
calculated from the normalizations for the fitted X-ray spectral 
models, which are proportional to the volume emission measures.  
We assumed fully ionized H and He, with $n_e = 1.2 n_H$.
Results are 0.66, 0.15, and 0.30-0.55 (dependent
on geometry) electrons cm$^{-3}$
respectively.  The X-ray emission from the interior implies a pressure
of $\approx 1 \times 10^{-10}$ dyne cm$^{-2}$, four times that derived 
for the filaments.  The thermal energy of the hot interior gas is 
$\approx 3 \times 10^{50}$ ergs.  The observed sum of kinetic and 
thermal energy is therefore $\approx 5 \times 10^{50}$ ergs which is 
a lower limit for $E_0$.  

The ionization timescale for the diffuse interior, with the dynamical 
age calculated in \S6.1, would imply a lower density of 0.06--0.07 
cm$^{-3}$, rather than the higher value calculated from the X-ray
normalization above.  In part, this value may be an underestimate, 
as the optical velocity may not accurately represent the expansion
of the remnant.  If, for example, we took the X-ray emission as 
being generated behind the expanding shock, its temperature would
imply a shock speed of 540 km s$^{-1}$ and (if approximately 
blast-wave expansion) an expansion speed of 400 km s$^{-1}$, which 
in turn would give a SNR age of about 22,000 yr.  This age, combined 
with the ionization timescale, gives a density of about 0.1 cm$^{-3}$,
closer to the emission-measure-derived value.

Conditions in the outer shell are not known.  Obviously there is no  
X-ray limb brightening around most of the remnant.  The shock has 
slowed to a point where material is shock-heated to temperatures 
of $T \leq  8 \times 10^5$ K (0.07 keV), the approximate temperature 
below which the thermal X-rays are too soft to penetrate the ISM 
\citep{S00}.  In the absence of a clear PWN contribution, much of 
the interior emission can be attributed to ``fossil" radiation - that 
is, emission from the low-density gas in the remnant interior
\citep{S85,RP98} that was heated during earlier, more energetic stages 
of the SNR's expansion.  The prominent optical filaments show that the
adiabatic phase is finishing and the remnant is now cooling.  The age 
derived from the expansion of the filamentary shell is compatible with 
this.  Estimates of the remnant parameters are summarized in
Table~\ref{tab:xprop}.  

\begin{figure}
\plotone{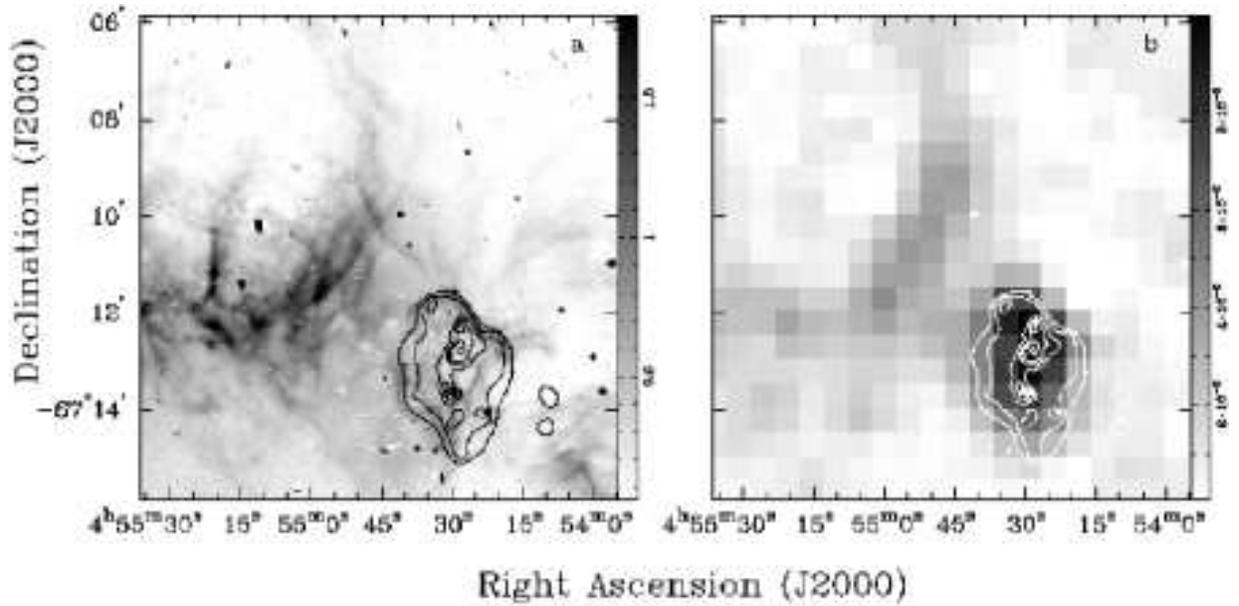}
\caption{(a) Adaptively smoothed ACIS X-ray contours overlaid
on a greyscale MCELS [\ion{O}{3}] image. (b) ACIS contours 
overlaid on a greyscale image of 843 MHz radio emission. Contours 
are scaled as in Fig.~\ref{fig:xray}.  The larger shell to the
northeast is from the neighboring N9 \ion{H}{2} region.}
\label{fig:lgrad}
\end{figure}

\begin{figure}
\epsscale{0.8}
\plotone{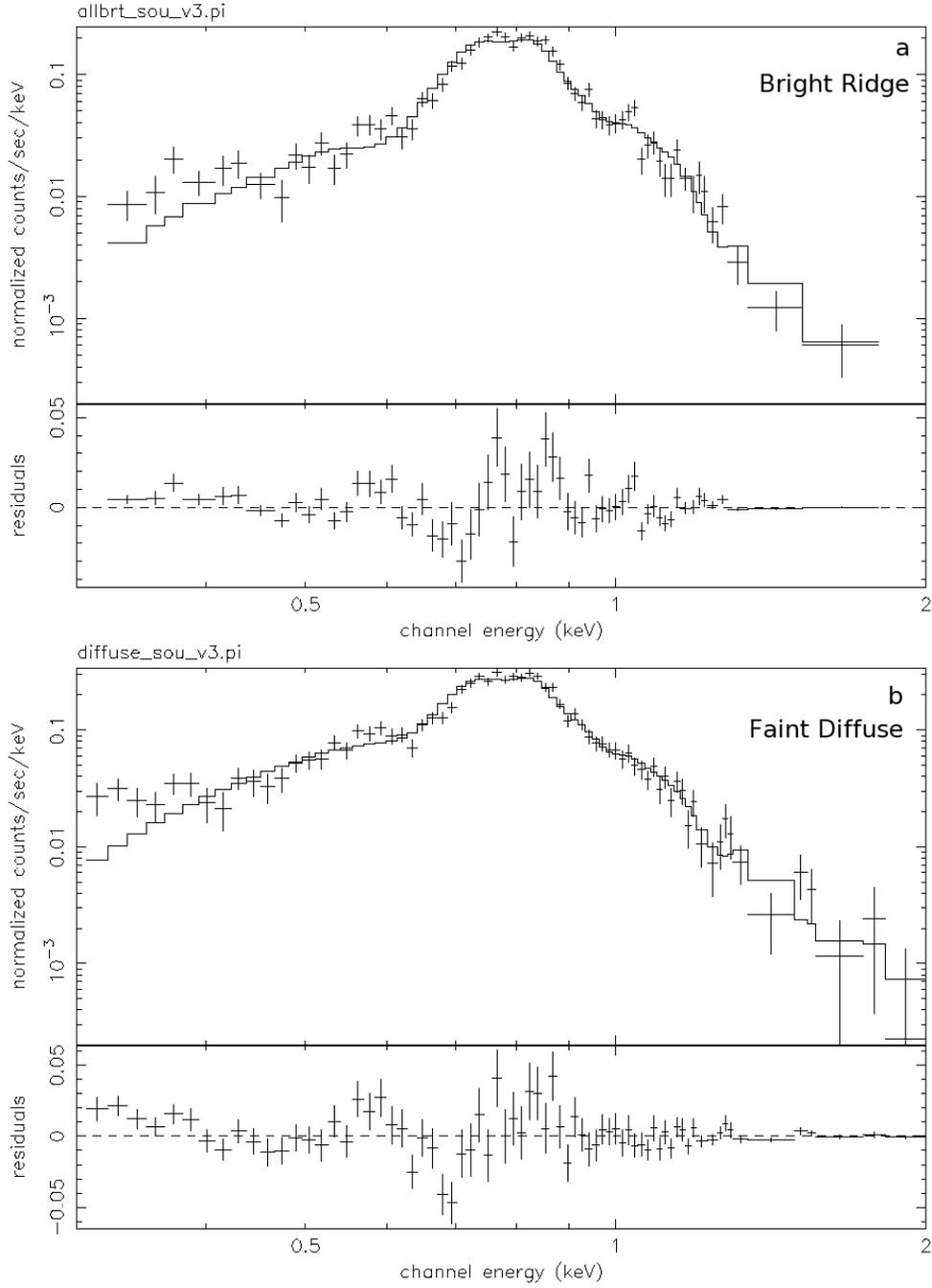}
\caption{ACIS spectra from (a) the Bright Ridge of central 
emission, (b) the Faint Diffuse emission from the rest of the
SNR.}
\label{fig:spec}
\end{figure}

\begin{figure}
\epsscale{0.8}
\plotone{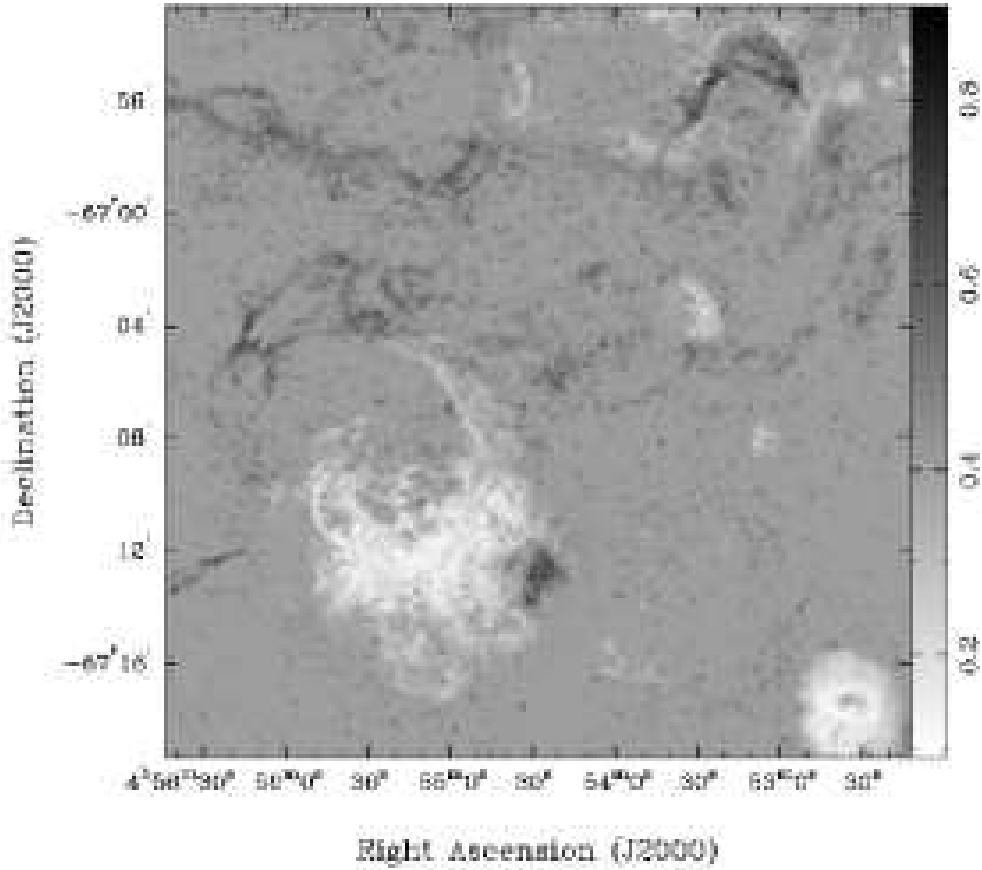}
\caption{[\ion{S}{2}]/H$\alpha$] map from flux-calibrated MCELS
images. Input images were clipped at 2$\sigma$ of background.
The final image has been smoothed by a Gaussian with $\sigma=2$.
The sky background level has been set to 0.4, so shocked gas 
([\ion{S}{2}]/H$\alpha > 0.4$) appears in black and photoionized
gas ([\ion{S}{2}]/H$\alpha < 0.4$) appears in white. Some noise is 
present due to the residue from star subtractions. N9 is toward
the bottom of the image, while the neighboring remnant in N4 is 
seen to the upper right.}
\label{fig:s2ha}
\end{figure}

The domination of a remnant by ``fossil" radiation is one explanation
often extended for the existence of ``mixed morphology" SNRs,
which have shell-like radio but centrally concentrated thermal X-ray 
emission, and lack a compact central source \citep[e.g.,][]{RP98,C+99}.  
N9 appears to fulfil at least two of these conditions. The X-ray data 
do not indicate a CCO (\S5), and the radio spectrum for the SNR is not consistent with the presence of a PWN.  Also, its X-ray morphology (\S3)
shows more prominent X-ray emission in the center than at the SNR limb,
though the X-ray bright regions are distributed in a patchy, elongated
fashion.  On the other hand, it is difficult to determine a clear radio 
shell from the available data (Figure~\ref{fig:lgrad}b), which makes this
classification less certain.  However, one must also take into account 
the low resolution of the radio images, and confusion with thermal 
emission from the \ion{H}{2} region.

It is difficult to determine the kind of SN which produced this 
remnant. Because N9 is an \ion{H}{2} region, the stellar population is
expected to be young and any associated SN is expected to be Type II 
with consequent production of a neutron star; but there is no strong 
evidence for any compact object associated with SNR 0454$-$672.  There 
is certainly no energetic pulsar within the remnant and the nearest
unresolved X-ray source is 1\farcm2  from the center.  If created at 
the center, this would require a velocity of $\approx 600$ km s$^{-1}$ 
to reach its present location, which is not an unreasonable velocity 
for a young neutron star.  It is also not clear that a $3 \times 10^4$ 
year old CCO would still be detectable or that the collapse did not 
create a black hole.  Thus, we cannot rule out a Type II SN origin.
The interior, however, shows signs of enhanced Fe abundance, consistent 
with a Type Ia SN origin.  For example, the  O/Fe number ratio in the 
SNR, based on our fitted abundances, is $\leq1.3$, compared to a 
typical LMC ratio of 13.2 \citep[based on][]{RD92}.  Models for the 
nucleosynthetic yields from Type Ia SNe suggest O/Fe ratios $<1$, while 
for Type II SNe the ratio is generally $\ga10$ \citep{I+99}.  Although
the ratios in the SNR have large uncertainties and are affected by 
the presence of swept-up ISM, the relative contribution of iron 
appears to be significantly greater than one would expect from a
Type II SN origin. This evidence is not as strong as in the case of 
DEM~L71 \citep{H+03}, but DEM L71 is a younger and brighter remnant. 

\acknowledgments

We thank the staff of the {\it Chandra} X-ray Center for smooth acquisition
of the data and for initial data processing.  X-ray data analysis was
supported by CXC/NASA Grant GO3-4069A. RMW acknowledges additional 
support from NASA grant NNG05GC97G. We thank Anne Green and Vince
McIntyre for use of the 843 MHz data in Figure 5 and Bryan Dunne for 
getting the column density from the \citet{KS+03} survey.

\newpage


\begin{deluxetable}{lccccccccc}
\tabletypesize{\scriptsize}
\tablecaption{Best-fit spectral models to SNR regions}
\tablehead{
\colhead{Parameter} & 
\multicolumn{2}{c}{Whole SNR} &
\multicolumn{2}{c}{Bright Ridge} &
\multicolumn{2}{c}{Faint Diffuse} }
\startdata
component  & vpshock & vmekal & vpshock & vmekal & vpshock & vmekal \\[6pt]
kT (keV) & 0.34$^{+0.02}_{-0.01}$ & 0.206$^{+0.012}_{-0.009}$ & 
	0.33$^{+0.04}_{-0.02}$ & 0.21$^{+0.02}_{-0.01}$ & 
	0.34$^{+0.03}_{-0.01}$ & 0.20$^{+0.01}_{-0.01}$ \\[6pt]
 O/O$_\sun$ & 0.04$^{+0.01}_{-0.01}$ & 0.004 ($< 0.01$) & 
 	0.01 ($< 0.04$) & 0 ($< 0.008$)  & 
 	0.05$^{+0.02}_{-0.02}$ & 0.012$^{+0.014}_{-0.011}$  \\[6pt]
 Ne/Ne$_\sun$ & 0.42$^{+0.08}_{-0.07}$ & 0.29$^{+0.08}_{-0.06}$ & 
 	0.6$^{+0.2}_{-0.2}$  & 0.5$^{+0.2}_{-0.2}$ &
 	0.37$^{+0.09}_{-0.08}$ & 0.26$^{+0.09}_{-0.08}$ \\[6pt]
 Fe/Fe$_\sun$ & 1.4$^{+0.2}_{-0.1}$ & 0.8$^{+0.1}_{-0.2}$ &
 	2.1$^{+0.6}_{-0.2}$ & 1.2$^{+0.5}_{-0.2}$  &
 	1.2$^{+0.2}_{-0.2}$ &  0.7$^{+0.2}_{-0.2}$ \\[6pt]
$\tau$ (cm$^{-3}$ s) & 8$^{+3}_{-2} \times10^{10}$ & \nodata &
	1.1$^{+0.7}_{-0.5} \times10^{11}$ &  \nodata &
	7$^{+5}_{-2} \times10^{10}$ &  \nodata  \\[6pt]
norm\tablenotemark{a} (cm$^{-5}$) &  1.0$^{+0.3}_{-0.2} \times10^{-3}$ & 
	4.7$^{+0.6}_{-0.5} \times10^{-3}$ &
	2.7$^{+0.8}_{-0.9} \times10^{-4}$ & 
	1.0$^{+0.2}_{-0.2} \times10^{-3}$ &
	7$^{+2}_{-2} \times10^{-4}$ & 
	3.4$^{+0.5}_{-0.2} \times10^{-3}$ \\[6pt]
$\chi^2_{\rm red}$ & 1.48 & 1.88 & 1.72 & 2.26 &  1.21 & 1.44 \\
dof & 208 & 209 & 83 & 84 & 190 & 191 \\
\enddata
\tablenotetext{a}{The normalization for these models is equal 
to the volume emission measure multiplied by $10^{-14}/(4\pi D^2)$,
where $D$ is the distance to the source in cm.}
\tablecomments{N$_{\rm H}$ is fixed to a value of $2\times10^{21}$ 
cm$^{-2}$ from H I measurements for this pointing in the LMC
\citep{KS+03} and the Galaxy \citep{DL90}. Spectra cover the range 
between 0.3-8.0 keV.  Elements not listed are fixed to a LMC mean 
abundance of 30\% solar \citep{RD92}.  Quoted errors are the 
statistical errors in each fit parameter at the 90\% uncertainty level.}
\label{tab:spec}
\end{deluxetable}


\begin{deluxetable}{lcccccc}
\tablecaption{X-ray Volume Emission Measures}
\tablehead{
\colhead{Region} & 
\colhead{Vol. EM (cm$^{-3}$)} & 
\colhead{Volume (cm$^{3}$)} 
 }
\startdata
Northern bright & $2.0\times10^{57}$ & $7.9 \times10^{57}$ \\
		&			& $2.7\times10^{58}$ \\
Central bright &  $4.1\times10^{57}$ & $1.1\times10^{58}$ \\
Southern bright &  $1.6\times10^{57}$ & $5.7\times10^{57}$ \\
Faint diffuse &  $2.1\times10^{58}$ & $1.1\times10^{60}$ \\
\enddata
\label{tab:xvolem}
\end{deluxetable}


\begin{deluxetable}{lcccccc}
\tablecaption{Point Sources}
\tablehead{
\colhead{RA(2000)} & 
\colhead{Dec (2000)} &
\colhead{Counts (0.4-7 keV)} &
\colhead{Counterpart\tablenotemark{a}} &
\colhead{Comment} }
\startdata
04 54 37.41 & -67 14 48.4 & 41 & none     & \nodata \\
04 54 33.42 & -67 14 50.9 & 18 & yes      & \nodata \\
04 54 29.20 & -67 11 34.1 & 15 & none?    & \nodata \\
04 54 17.16 & -67 13 13.4 & 22 & none     & inside remnant \\
04 54 15.43 & -67 15 50.9\tablenotemark{b} & 15 & none     & \nodata \\
04 54 00.14 & -67 12 54.6\tablenotemark{b} & 81 & none?    & hard spectrum \\
04 53 58.00 & -67 13 35.8\tablenotemark{b} & 61 &yes, bright& soft spectrum \\
\enddata
\small
\tablenotetext{a}{Optical counterparts identified in Digital Sky 
Survey and MCELS ``continuum" images.}
\tablenotetext{b}{Outside of field covered by Figure 1.} 
\normalsize
\label{tab:src}
\end{deluxetable}


\begin{deluxetable}{lcccccc}
\tablecaption{Physical Properties of SNR 0454$-$672}
\tablehead{
\colhead{Parameter} & 
\colhead{Value\tablenotemark{a}} }
\startdata
X-ray luminosity (erg s$^{-1}$) & $5.7 \times 10^{35}$  \\
Explosion energy (erg) & $ \geq 4 \times 10^{50}$  \\
Age (yr)  & $3 \times 10^4$  \\
Mass ($M_{\odot}$) & 1300 in shell \\
& 160 in interior hot gas \\
& 7 in central 9 pc \\
\enddata
\small
\tablenotetext{a}{Assuming a LMC distance of 50 kpc \citep{W90}}
\normalsize
\label{tab:xprop}
\end{deluxetable}

\newpage

\begin{thebibliography}{}
\bibitem[Bandiera \& Petruk(2004)]{BP04}Bandiera, R. \& Petruk, O.\ 2004, \aap, 419, 419
\bibitem[Chu(1997)]{C97}Chu, Y.-H.\ 1997, \aj, 113, 1815
\bibitem[Chu \& Kennicutt(1988)]{CK88}Chu, Y.-H. \& Kennicutt, R.C.\ 1988, \aj, 96, 1874
\bibitem[Cox(1972)]{C72}Cox, D.P.\ 1972, \apj, 178, 159
\bibitem[Cox et al.(1999)]{C+99} Cox, D.~P., Shelton, R.~L., 
Maciejewski, W., Smith, R.~K., Plewa, T., Pawl, A., \& R{\'o}{\.z}yczka, 
M.\ 1999, \apj, 524, 179 
\bibitem[Dickel(1991)]{D91}Dickel, J.\ 1991, in Woosley, S. ed.,
Supernovae, The tenth Santa Cruz Summer workshop in Astronomy and
Astrophysics, (Berlin: Springer) 675
\bibitem[Dickel et al.(2005)]{D+05}Dickel, J., McIntyre, V., Gruendl,
R., \& Milne, D.\ 2005, \aj, in press, February
\bibitem[Dickey \& Lockman(1990)]{DL90}Dickey, J.~M., \& 
Lockman, F.~J.\ 1990, \araa, 28, 215 
\bibitem[Evans et al.(1980)]{E+80}Evans, W.~D. et al.\ 1980, \apj, 237, L7
\bibitem[Filipovic et al.(1998)]{F+98}Filipovic, M.~D. et al.\ 1998, \aaps, 127, 119
\bibitem[Gaensler et al.(2003)]{G+03}Gaensler, B.~M., Hendrick, S.~P., Reynolds, S.~P., \& Borkowski, K.~J.\ 2003, \apj, 594, L111
\bibitem[Gotthelf et al.(2005)]{G+05} Gotthelf, E.~V., 
Halpern, J.~P., \& Seward, F.~D.\ 2005, \apj, 627, 390 
\bibitem[Haynes et al.(1991)]{H+91}Haynes, R., Klein, U.,Wayte, S., Wielebinski, R., Murray, J., Bajaja, E., Meinert, D., Buczilowski, U., Harnett, J., Hunt, A., Wark, R., \& Sciacca, L.\ 1991, \aap, 252, 475
\bibitem[Hendrick et al.(2003)]{HBR03}Hendrick, S., Borkowski, K., \& Reynolds, S.\ 2003,  \apj, 593, 370
\bibitem[Hughes et al.(1995)]{H+95}Hughes, J.P. et al.\ 1995,  \apj, 444, L81
\bibitem[Hughes et al.(1998)]{H+98}Hughes, J.P., Hayashi, I., \& Koyama, K.\ 1998, \apj, 505, 732
\bibitem[Hughes et al.(2003)]{H+03}Hughes, J.P. Ghavamian, P., Rakowski, C.E., \& Slane, P.O.\ 2003,  \apj, 582, L95
\bibitem[Iwamoto et al.(1999)]{I+99}Iwamoto, K., Brachwitz, F., Nomoto, K., Kishimoto, N., Umeda, H., Hix, W. R., \& Thielemann, F.\ 1999, \apjs, 125, 439
\bibitem[Kaaret et al.(2001)]{K+01}Kaaret et al.\ 2001, \apj, 546, 1159
\bibitem[Kaspi \& Helfand(2002)]{KH02}Kaspi, V. \& Helfand, D.~J.\ 2002, astro-ph/0201183
\bibitem[Inoue et al.(1983)]{IKT83}Inoue, H., Koyama, K., Tanaka, Y.\  1983, in SNR and Their Emission, ed J. Danziger \& P. Gorenstein, 535
\bibitem[Kim et al.(2003)]{KS+03} Kim, S., Staveley-Smith, L., Dopita, M.~A., Sault, R.~J., Freeman, K.~C., Lee, Y., \& Chu, Y.-H. 2003, \apjs, 148, 473
\bibitem[Mazets et al.(1979)]{M+79}Mazets, E. P., Golentskii, S. V., Ilinskii, V.~N., Aptekar, R.~L., \& Guryan, I.~A.\ 1979, Nature, 282, 587
\bibitem[Mathewson et al.(1984)]{M+83}Mathewson, D.~S. et al.\ 1983, \apjs, 51, 345
\bibitem[Osterbrock(1989)]{O89}Osterbrock, C.\ 1989, {\it Astrophysics of Gaseous Nebulae and Active Galactic Nuclei}, (Mill Valley: University Science Books), Table 4.1
\bibitem[Rho \& Petre(1998)]{RP98}Rho, J. \& Petre, R.\ 1998, \apj, 503, L167
\bibitem[Rothschild, Kulkarni, \& Lingenfelter(1994)]{R+94}Rothschild, R.~E, Kulkarni, S.~R., \& Lingenfelter, R.~E.\ 1994, Nature, 368, 432
\bibitem[Russel \& Dopita(1992)]{RD92} Russel, S.~C. \& Dopita, M.~A. 1992, 
 \apj, 384, 508
\bibitem[Sedov(1959)]{S59} Sedov, L.~I.\ 1959, {\it Similarity and Dimensional Methods in Mechanics} (New York: Academic)
\bibitem[Seward(1985)]{S85} Seward, F.~D.\ 1985, Comments Astrophys. XI, 1, 15
\bibitem[Seward(2000)]{S00} Seward, F.~D.\ 2000, in Allen's Astrophysical Quanties, ed. Arthur N. Cox (Springer, New York), pp. 193,195
\bibitem[Seward  et al.(2003)]{S+03}Seward, F.~D., Slane, P., Smith, R., and Sun, M.\ 2003, \apj, 584, 414
\bibitem[Smith et al.(1994)]{S+94}Smith, R.~C., Chu, Y.-H., Mac Low, M.-M., Oey, M.~S. \& Klein, U.\ 1994, \aj, 108, 1266
\bibitem[Smith(2002)]{S02}Smith, R.C., 2002, \url{http://ctio.noao.edu/$\sim$mcels/snrs/snrcat.html }
\bibitem[Smith et al.(1999)]{S+99}Smith, R.~C. \& The MCELS Team 1999, in IAU Symp. 190, New Views of the Magellanic Clouds, ed. Y.-H. Chu et al. (San Francisco: ASP), 28
\bibitem[Wang et al.(2001)]{W+01}Wang, Q.~D., Gotthelf, E.~V., Chu, Y.-H., \& Dickel, J.~R.\ 2001, \apj, 559 275
\bibitem[Westerlund(1990)]{W90}Westerlund, B.~E.\ 1990, \aapr, 2, 29
\bibitem[Williams et al.(2000)]{W+00}Williams, R. et al.\ 2000,  \apj, 536, L27
\bibitem[Williams et al.(1999)]{W+99}Williams, R. et al.\ 1999,  \apjs, 123, 467
\bibitem[Williams et al.(2005)]{W+05} Williams, R.~M., Chu, Y.-H., Dickel, J.~R.,  Gruendl, R.~A., Seward, F.~D., Guerrero, M.~A. \& Hobbs, G.\ 2005,  \apj, 628, 704, astro-ph/0504609
\end{thebibliography}
\end{document}